\documentclass[12pt]{article}

\usepackage{amsfonts}
\newlength{\abstractwidth}
\usepackage{epsf}
\usepackage{color}
\usepackage{graphicx}

\renewcommand{\thefootnote}{\fnsymbol{footnote}}
\renewcommand{\thanks}[1]{\footnote{#1}}
\newcommand{\starttext}{
\setcounter{footnote}{0}
\renewcommand{\thefootnote}{\arabic{footnote}}}

\newcommand{\bea}{\begin{eqnarray}}
\newcommand{\eea}{\end{eqnarray}}
\newcommand{\ee}{\end{equation}}
\newcommand{\be}{\begin{equation}}

\def\cN{{\cal N}}
\def\cO{{\cal O}}

\def\bC{{\bf C}}

\def\bR{{\bf R}}

\def\tr{{\rm tr}}

\def\half{ {1\over 2}}
\def\p{\partial}

\def\ep{\varepsilon}

\def\g{\gamma}

\def\ch{{\rm \, ch }}
\def\sh{{\rm \, sh }}

\def\bw{{\bar w}}

\def\no{\nonumber}
\def\sm{\smallskip}

\long\def\symbolfootnote[#1]#2{\begingroup%
\def\thefootnote{\fnsymbol{footnote}}\footnote[#1]{#2}\endgroup}

\begin{document}
\starttext
\setcounter{footnote}{0}

\begin{flushright}
UCLA/09/TEP/43 \\
CPHT-RR025.0409 \\
21 April  2009
\end{flushright}

\bigskip

\begin{center}

{\Large \bf Janus solutions in M-theory\symbolfootnote[2]{\noindent This work was
supported in part by NSF grant PHY-07-57702.}

} \medskip
 
\medskip

\vskip .4in

{\large  Eric D'Hoker$^{a}$, John Estes$^{b}$, Michael Gutperle$^{a}$,
and  Darya Krym$^{a}$}

\vskip .2in

{$\ ^{a}$ \sl Department of Physics and Astronomy }\\
{\sl University of California, Los Angeles, CA 90095, USA}\\
{\tt \small dhoker@physics.ucla.edu; gutperle@physics.ucla.edu;
daryakrym@gmail.com}

\vskip .2in

{$\ ^{b}$\sl  Centre de Physique Th«eorique, Ecole Polytechnique,}\\
{\sl FÐ91128 Palaiseau, France}\\
{\tt  \small johnaldonestes@gmail.com}
\end{center}

\vskip .2in

\begin{abstract}

\vskip 0.1in

We present a one-parameter deformation of the $AdS_{4}\times S^{7}$ 
vacuum, which is a regular solution in M-theory, invariant under 
$SO(2,2)\times SO(4) \times SO(4)$, and which preserves 16 supersymmetries. 
The solution corresponds to a holographic realization of a Janus-like interface/defect theory, despite the absence of a dilaton in  M-theory. The 2+1-dimensional CFT 
dual results from the maximally symmetric CFT through the insertion of a 
dimension 2 operator which is localized along a 1+1-dimensional linear 
interface/defect, thereby partially breaking the superconformal symmetry. 
The solution admits a regular ABJM reduction to a quotient solution which is 
invariant under $SO(2,2) \times SO(4)\times U(1)^2$, preserves 12 
supersymmetries, and provides a Janus-like interface/defect  solution in 
ABJM theory. 
 
\end{abstract}

%\newpage

%\tableofcontents

\newpage

\baselineskip=16pt
\setcounter{equation}{0}
\setcounter{footnote}{0}

\section{Introduction}

The Janus solution in Type IIB supergravity \cite{Bak:2003jk} provides one of the 
simplest deformations of the maximally supersymmetric $AdS_{5}\times S^{5}$ solution. 
The original Janus solution has vanishing 3-form fields and breaks all supersymmetries, 
but solutions with non-vanishing 3-form fields which preserve various degrees of 
supersymmetry exist as well. Janus solutions with 4 and 16 supersymmetries were 
constructed respectively in 
\cite{D'Hoker:2006uu,Clark:2005te}, and in 
\cite{D'Hoker:2007xy,D'Hoker:2007xz}\footnote{See  
\cite{Yamaguchi:2006te,Gomis:2006cu,Lunin:2006xr,Lunin:2007ab} 
for earlier and related work on the holographic description of BPS defects.}. 
The Type IIB Janus solutions exhibit a common characteristic, namely 
their non-trivial space-dependence of the dilaton field, and smooth interpolation 
between several different asymptotic $AdS_5 \times S^5$ regions, each of which 
has an independent constant dilaton expectation value.
The holographic duals to these Janus solutions are interface/defect theories in 
which the gauge coupling is constant throughout the bulk of each half-space,
but is allowed to jump across a planar 2+1-dimensional interface/defect,
where the half-spaces join together. Generally, local gauge invariant operators  
which are localized on the interface/defect may be inserted, and these are 
in fact required for supersymmetry \cite{Clark:2004sb,D'Hoker:2006uv}.

\sm

At first sight, it may seem that no Janus solutions can exist  in M-theory,
since there is no dilaton field in M-theory. 

\sm

In this paper we will show that, contrary to this naive expectation, there actually exists 
a regular one-parameter family of deformations of the maximally symmetric 
$AdS_4 \times S^7$ solution for M-theory, which is invariant under 
$SO(2,2) \times SO(4) \times SO(4)$, preserves 16 supersymmetries, and can 
be interpreted as a Janus-like solution in M-theory. Its full supergroup invariance 
is $OSp(4|2,\bR) \times OSp(4|2,\bR)$. This family of solutions is characterized 
not by the space-dependence of the dilaton -- since there is no dilaton in M-theory 
-- but rather by the space-dependence of the 4-form field of M-theory. 
Holographically, these fields correspond to adding counterterms which are 
localized at the interface/defect.  As we shall show below, the Janus-like deformation 
in M-theory corresponds to the addition of such counterterms for the M2-brane CFT. 

\sm

The remainder of this paper is organized as follows. In section 2, we  begin by 
carrying out a linearized analysis in an $AdS_4$ background for the effect of 
inserting a dimension 2 operator on the interface/defect. We shall confirm that smooth 
deformations exist at this order. In section 3, we exhibit the exact one-parameter 
family of Janus-like solutions in M-theory, and discuss their holographic interpretation.
In section 4, we show that our Janus solutions are invariant under the $Z_k$
transformations required to carry out the Aharony, Bergman, Jafferis, Maldacena 
(ABJM)  \cite{Aharony:2008ug} reduction of $AdS_4 \times S^7$ to $AdS_4 \times CP_3$ in Type IIA. 
Quotienting our Janus deformations of $AdS_4 \times S^7$
by $Z_k$, $k \not= 1,2$, yields new solutions which are invariant under 
$SO(2,2) \times SO(4) \times U(1)^2$, preserve 12 supersymmetries, and 
exhibit invariance under the supergroup $OSp(3|2,\bR) \times OSp(3|2,\bR)$.

%%%%%%%%%%%%%%%%%%%%%%%%%%%%%%%%%%%%%%%%%%%%%%%
%%%%%%%%%%%%%%%%%%%%%%%%%%%%%%%%%%%%%%%%%%%%%%%
\section{Linearized analysis}
\label{linana}
\setcounter{equation}{0}
%%%%%%%%%%%%%%%%%%%%%%%%%%%%%%%%%%%%%%%%%%%%%%%
%%%%%%%%%%%%%%%%%%%%%%%%%%%%%%%%%%%%%%%%%%%%%%%

The starting point of the linearized analysis is 4-dimensional gravity  with 
negative cosmological constant,  minimally coupled to a scalar or pseudo-scalar 
field $\phi$ of mass $m$. 
In the following we recall some basic features of the AdS/CFT correspondence 
\cite{Maldacena:1997re,Gubser:1998bc,Witten:1998qj} (for reviews, see 
\cite{Aharony:1999ti,D'Hoker:2002aw}).
The action is given by,
\be
\label{startact}
S= \int d^{4} x  \sqrt{-g}\Big( R-\Lambda
 -{1\over 2} \partial_{\mu} \phi \partial^{\mu} \phi-{1\over 2} m^{2} \phi^{2}\Big)
\ee
The Janus solution \cite{Bak:2003jk} provides a simple holographic description 
of an interface conformal field theory. Following this example, we parametrize
the $AdS_4$ metric by a slicing coordinate $\mu$, and a transverse 
$AdS_{3}$ space for each value of $\mu$, and choose  $\phi$ to only 
depend on $\mu$,
\bea
ds^{2}= f(\mu) \Big( d\mu^{2}+ ds_{AdS_{3}}^{2}\Big)
\hskip 1in 
 \phi=\phi(\mu) 
\eea
We set the cosmological constant equal to $\Lambda=-6$,  
so that the $AdS_{4}$ space  has unit radius.   
The pure $AdS_{4 }$ solution is then given by, 
\be
\label{adsback}
f(\mu)={1\over \cos^{2}(\mu)} 
\hskip 1in 
\phi=0
\ee
where the range of $\mu$ is $\mu \in [-\pi/2,\pi/2]$.

\sm

To obtain a Janus-like deformation, at linearized order, we choose the mass-square  
of  the field $\phi$ to be $m^{2}=-2$.  To justify this choice, we recall the 
AdS/CFT relation between $m^2$ and the conformal dimension $\Delta$ of the 
operator which is dual to $\phi$ in the dual CFT, 
\be
\label{massdrel}
m^{2}= \Delta(\Delta-3)
\ee
For $m^2=-2$, this equation has two solutions, namely $\Delta =1$, and $\Delta=2$.
As a result, near the boundary components  of $AdS_{4}$ at $\mu\sim \pm \pi/2$, the  
asymptotic behavior of  $\phi$ is given by,
\be
\label{phibound}
\phi (\mu) = \phi_{1} (\mu \mp \pi/2)  +\phi_{2} (\mu \mp \pi/2)^{2} + \cdots
\ee 
where $\phi_1$ and $\phi_2$ are constants.
Note that the mass $m^{2}=-2$ lies in the range $-9/4 < m^{2} <-5/4$ where 
both the modes associated with $\phi_{1}$ and $\phi_{2}$ in (\ref{phibound}) 
are normalizable. Hence, there is an ambiguity in identifying which  
mode sources the operator and which mode turns on the expectation value. 
This ambiguity is related to the fact that the equation (\ref{massdrel}) for $m^{2}=-2$ 
has two solutions corresponding to operators of conformal dimension $\Delta=1$ 
and $\Delta=2$. Generally, in unitary CFTs,  both scaling dimensions  are  allowed. 
The ambiguity can be resolved, however,  for supergravity theories 
\cite{Breitenlohner:1982jf,Mezincescu:1984ev}. For the case at hand, the results of these 
papers imply that pseudo-scalar fields of mass $m^{2}=-2$ correspond to 
operators with $\Delta=2$, for which the mode $\phi_{1}$ sources the operator, 
while $\phi_{2}$ turns on its expectation value.  Scalar fields of mass 
$m^{2}=-2$ correspond to operators with $\Delta=1$, for which the roles of $\phi_{1}$ 
and $\phi_{2}$ are reversed.

\sm

In the remainder of the paper, we will focus on the first case. 
Thus, $\phi$ will be a pseudo-scalar, and we shall denote the CFT dual 
operator by $\cO_2$; its dimension is $\Delta=2$. The operator $\cO_2$ would  be
sourced -- in the bulk of the CFT -- by the $\phi_{1}$ mode. We shall set 
$\phi_1=0$, so that the operator is  {\sl not sourced} in the bulk of the CFT.  
Thus, in the gravity dual, $\cO_2$ will not be sourced in the two 
boundary half-spaces that are dual to the bulk of the CFT.
The mode $\phi_2$ then corresponds to the vacuum expectation value of 
$\cO_2$ in the bulk of the CFT. The operator $\cO_2$ will be sourced 
on the interface/defect, however, confirming that its conformal dimension 2 
is precisely the one needed to maintain conformal invariance on 
the interface/defect.

\sm 

In the following, we consider the linearized problem of small fluctuations around 
the $AdS_{4}$ background (\ref{adsback}). The equation for the pseudo-scalar 
field $\phi$ becomes,
\be
{d^{2} \phi\over d\mu^{2}}+ 2 \tan\mu {d\phi\over d\mu}+ {2\over \cos^{2}\mu} \phi =0
\ee
which can be solved   exactly, 
\be
\label{phisola}
\phi(\mu)= \phi_{1}  \cos \mu \sin\mu+\phi_{2} \cos^{2}\mu
\ee
Upon setting $\phi_1=0$, as was advocated above, the linearized solution $\phi(\mu)$
of (\ref{phisola}) indeed reproduces the asymptotic behavior of (\ref{phibound}).  
Note that both the metric back-reaction as well as the coupling of the pseudo-scalar $\phi$ 
to other fields will be of order $(\phi_{2})^{2}$ and can be neglected in the linearized approximation.    
The solution of the full non-linear equation  can only be obtained numerically and will 
not be needed in this paper, since an exact Janus solution of 11-dimensional 
supergravity  will be presented  in section 3.

%%%%%%%%%%%%%%%%%%%%%%%%%%%%%%%%%%%%%%%%%%%%%%%
%%%%%%%%%%%%%%%%%%%%%%%%%%%%%%%%%%%%%%%%%%%%%%%
\subsection{Holographic interpretation}
 %%%%%%%%%%%%%%%%%%%%%%%%%%%%%%%%%%%%%%%%%%%%%%%
%%%%%%%%%%%%%%%%%%%%%%%%%%%%%%%%%%%%%%%%%%%%%%%

In this section we adapt an argument given in \cite{D'Hoker:2007xy} to  the context 
of the $AdS_{4}/CFT_{3}$ correspondence  in M-theory. Our goal is to confirm that the 
linearized solution presented in the previous section indeed corresponds to sourcing the 
operator $\cO_2$ dual to the field $\phi$ on a 1+1-dimensional interface/defect, instead 
of sourcing $\cO_2$ on the entire 2+1-dimensional boundary.
The metric for the $AdS_{3}$ slicing of $AdS_{4}$ takes the following form near  
$\mu= \pm \pi/2 $. 
\bea
ds^{2}
&=&
{1\over \cos^{2} \mu}\Big(d\mu^{2}+ {1\over z^{2}} (-dt^{2}+dx^{2}+dz^{2})\Big)
\no\\
&=& 
{1\over z^{2} \cos^{2}\mu}\Big( z^{2}d\mu^{2}+  (-dt^{2}+dx^{2}+dz^{2})\Big)
\eea
It is now easy to see that the boundary of $AdS_{4}$ in this coordinate system consists 
of  three components \cite{Bak:2003jk}.  The limits  $\mu\to {\pm}\pi/2 $  correspond to 
two 2+1-dimensional half spaces (since the range of $z$ is the positive real numbers). 
The two half spaces are glued together along a 1+1-dimensional interface/defect 
which is associated with the boundary of the $AdS_{3}$ slice and is reached as $z\to 0$ with $\mu$ arbitrary.
With the definition $\epsilon= |\cos(\mu)| z$, the boundary components
  are   reached uniformly as $\epsilon\to 0$. The boundary source for an operator ${\cal O}_{2}$ can be obtained by isolating the term in $\phi$ which scales like $\epsilon^{3-\Delta}$.
\bea
\phi_{source} &=& \lim_{\epsilon\to 0}  \left ( \epsilon^{\Delta-3} \phi(\mu) \right )
\no\\
&=&\lim_{  \epsilon \to 0} \left ( {|\cos (\mu) | \over z}  \phi_{2} \right )
\label{nonnorma}
\eea
The limits depends upon the direction along which $\ep =0$ is being approached.
As one keeps $z\not=0$ fixed, and takes $\mu \to \pm \pi/2$, the two half spaces 
$\mu = \pm \pi/2$ are approached away from the interface/defect. It follows  from 
(\ref{nonnorma}) that this limit, and thus the source for the ${\cal O}_{2}$ operator, 
vanishes.  
This means the dual operator ${\cal O}_{2}$ is not inserted in the boundary CFT 
away from the interface/defect.

\sm

The interface/defect is approached as one takes  $z\to 0$  with $\mu \neq \pm \pi/2$.  
In this case the limit in  (\ref{nonnorma}), and thus the source for the $\cO_2$ 
operator, diverges. This behavior indicates  the presence of a Dirac $\delta$-function source for the operator $\cO_2$ on the  1+1-dimensional interface/defect. 
This may be established directly by integrating $\phi_{source}$ over a small 
disk around $\epsilon =0$ in the $z, \mu$-plane. The corresponding  integral is 
given by $ \int d\mu \; dz \, z\;\phi_{source}$ and  is finite.  Its interpretation is 
that  a term which has $\delta$-function support on the interface/defect is 
being added to the Lagrangian of the 2+1-dimensional CFT,
\be
{\cal L}={\cal L}_{CFT_{3}}+ \lambda \; \delta(x^{\perp}) {\cal O}_{2}
\ee
where $x^\perp$ is the coordinate transverse to the 1+1-dimensional interface/defect.
The linearized analysis corresponds to a small perturbation with $\lambda \ll 1$. 
Since the 
conformal dimension of the operator ${\cal O}_{2}$  is two, its addition to
${\cal L}_{CFT_{3}}$ preserves  the (global) 1+1-dimensional conformal symmetry of the 
interface/defect, but breaks the full 2+1-dimensional conformal symmetry of the CFT.
 
%%%%%%%%%%%%%%%%%%%%%%%%%%%%%%%%%%%%%%%%%%%%%%%
%%%%%%%%%%%%%%%%%%%%%%%%%%%%%%%%%%%%%%%%%%%%%%%
\subsection{$\cN=8$ gauged supergravity}
 %%%%%%%%%%%%%%%%%%%%%%%%%%%%%%%%%%%%%%%%%%%%%%%
%%%%%%%%%%%%%%%%%%%%%%%%%%%%%%%%%%%%%%%%%%%%%%%

In this subsection, we shall inspect the supergravity fields on $AdS_4 \times S^7$
and identify viable candidates for the pseudo-scalar deformations studied above.
The spectrum of the Kaluza-Klein (KK) compactification of M-theory on 
$AdS_{4}\times S^{7}$ has been obtained by 
\cite{Biran:1983iy,Gunaydin:1985tc}\footnote{See 
\cite{Halyo:1998mc} for a dictionary between supergravity and AdS/CFT conventions.}. One gets infinite towers of KK-states organized  in representations of the  
$SO(8)$ R-symmetry group, as collected in Table 1.  
\begin{table}[htdp]
\begin{center}
\begin{tabular}{|c|c|c|c|}
\hline
spin&Dynkin label& $m^{2}$ &$\Delta$\\
\hline
2&$[n,0,0,0]$& $ 1/4 \, n(n+6)$ & $1/2  (n +6)$\\
\hline
1&$[n,1,0,0]+ [n-1,0,1,1]+[n-2,1,0,0]$& $1/4 \, n(n+2)$ & $1/2 (n+4)$\\
\hline
$0^{+}$&$[n+2,0,0,0]+[n-2,2,0,0]+[n-2,0,0,0]$ & $1/4(n+2)(n-4)$ & $1/2(n+2) $\\
\hline
$0^{-}$&$[n,0,2,0]+[n-2,0,0,2]$& $1/4( n(n+2)-8)$& $1/2(n+4)$\\
\hline
\end{tabular}
\caption{KK towers of bosonic fields from the supergravity multiplet for $n=0,1,2,\cdots$. 
Representations with negative Dynkin-labels are to be omitted. 
Here, $m^{2}$ is the mass of the supergravity field, and $\Delta$ is the dimension 
of its dual CFT operator. }
\end{center}
\label{default}
\end{table}%

There exists a consistent truncation of the theory to the ``massless'' $\cN=8$ multiplet 
which produces $\cN=8$ gauged supergravity in four dimensions and is given by 
the representations with $n=0$, summarized in Table 2.

\begin{table}[htdp]
\begin{center}
\begin{tabular}{|c|c|c|c|}
\hline
spin&$SO(8)$ representation& mass $m^{2}$ & dimensions $\Delta$ \\
\hline
2& ${\bf 1}$  & $ 0 $ & $3 $ \\
\hline
1&${\bf 28} $ & $ 0 $ & $2 $  \\
\hline
$\;\,0^{+}$&${\bf 35_{v} }$ & $-2$ & $1  $ \\
\hline
$\;\,0^{-}$&$ {\bf 35_{c}}$& $-2$& $2$ \\
\hline
\end{tabular}
\caption{The bosonic fields of 4-dimensional  $\cN=8$ gauged supergravity 
with the mass of the fields and the dimension of the dual operator in the CFT. }
\end{center}
\label{default}
\end{table}%
 
 Note that the 70 scalars of gauged supergravity split into 35 scalars (denoted 
 $0^{+}$ in Table 2)  and 35 pseudo-scalars scalars (denoted by $0^{-}$). 
 In the KK-reduction the scalars are obtained  from the reduction of the metric 
 component on the sphere whereas the pseudo-scalars are obtained from the 
 AST field strength on the sphere.  As discussed in section \ref{linana}, the 
 ambiguity for the conformal dimensions for the operators dual to the (pseudo-)
 scalars was resolved in \cite{Breitenlohner:1982jf,Mezincescu:1984ev} and 
 leads to the values displayed in Table 2.
 
\sm
 
In gauged $\cN=8$ supergravity a Janus interface/defect configuration can be 
obtained applying the linearized analysis of section \ref{linana} for a  pseudo-scalar 
transforming in the ${\bf 35_{c}}$ representation of the $SO(8)$ R-symmetry. 
This field should be dual to a dimension 2 operator in the $CFT_{3}$ defined by 
the decoupling limit of a large number of M2-branes. 
 
\sm

In principle one could try to solve the equations of motion for the $\cN=8$ gauged 
supergravity and a Janus Ansatz to obtain a fully non-linear solution dual to the 
insertion of such an operator.  Solving the full second order equations of motion is, 
however, prohibitively complicated. A different approach is to ask wether there 
are interface/defect solution which preserve some of the 32 supersymmetries 
and correspond to superconformal interface/defects.  Solving the resulting 
BPS-equation is generically  easier   than solving the equations of motion. 

\sm

Note that the  ${\bf 35_{c}}$   representation can be characterized as the 
rank four self-dual antisymmetric tensor representation of $SO(8)$. 
Turning on such a field will therefore 
break  the $SO(8)$ R-symmetry down to $SO(4)\times SO(4)$. Hence one  
expects the resulting theory to have  $SO(4)\times SO(4)$ unbroken R-smmetry.  

\sm

In the following we will use the results of \cite{D'Hoker:2008wc} 
to obtain a solution of 11-dimensional supergravity, which preserves sixteen 
supersymmetries, is locally asymptotic to $AdS_{4}\times S^{7}$, preserves 
$SO(2,2)\times SO(4)\times SO(4)$ symmetry and has all the characteristics of a 
fully non-linear and back-reacted  interface/defect solution discussed above.
   
%%%%%%%%%%%%%%%%%%%%%%%%%%%%%%%  
%%%%%%%%%%%%%%%%%%%%%%%%%%%%%%%  
 \section{The half-BPS Janus solutions in M-theory}
\setcounter{equation}{0}
%%%%%%%%%%%%%%%%%%%%%%%%%%%%%%%%
%%%%%%%%%%%%%%%%%%%%%%%%%%%%%%%  

The linearized analysis for a pseudo-scalar field $\phi$ which is dual to  a dimension 2
gauge invariant operator $\cO_2$ localized on a 1+1-dimensional interface/defect, 
and the inspection of multiplets for such fields in gauged $\cN=8$ supergravity on 
$AdS_4$, reveal a natural  candidate for an M-theory Janus solution.
The conformal invariance of the interface/defect theory, which is expected to be
maintained by the operator $\cO_2$, imposes $SO(2,2)$ symmetry. An expectation
value to an operator in the ${\bf 35_c}$ representation of $SO(8)$ leaves a 
residual $SO(4) \times SO(4)$ symmetry, as explained in the preceding section.
Thus, the full bosonic symmetry of the corresponding Janus solution in M-theory
should be $SO(2,2) \times SO(4) \times SO(4)$. Remarkably, such solutions
exist and preserve 16 supersymmetries, a result we shall establish in the 
present section.

\sm

The symmetry and supersymmetry conditions, advocated in the preceding 
paragraph, place the problem precisely in the context of the general analysis 
of half-BPS solutions in M-theory with $SO(2,2) \times SO(4) \times SO(4)$ 
symmetry, for which the general local exact solution was derived in 
\cite{D'Hoker:2008wc} (for space-times asymptotic
to either $AdS_4 \times S^7$ or $AdS_7 \times S^4$). 
We begin by briefly reviewing
the salient points of the solutions obtained in \cite{D'Hoker:2008wc}.  
The 11-dimensional metric Ansatz consists of a fibration of the unit
radius metric of $AdS_{3}\times S_2^{3}\times S_3^{3}$ over a 
2-dimensional Riemann surface $\Sigma$ with boundary $\p \Sigma$,
\bea
\label{metricansatza}
ds^2 = f_1^2 ds_{AdS_3}^2 + f_2^2 ds_{S_2^3}^2 
+ f_3^2 ds_{S_3^3}^2 +  ds_{\Sigma}^2
\eea
where $ds^2 _{AdS_3}$ and $ds^2 _{S_{2,3}^3}$ denote the metrics with unit 
radius on the corresponding spaces, which are invariant respectively under 
$SO(2,2)$ and $SO(4)$. The $SO(2,2) \times SO(4) \times SO(4)$-invariant 
Ansatz for the 3-form gauge potential 
$C_3$, and for the 4-form field strength $F_4=dC_3$ are given as follows,
\bea
\label{fluxansatz}
C_3 & = & 
b_1 \hat \omega_{AdS_{3}} + b_2 \hat \omega_{S_2^{3}} + b_3 \hat \omega_{S_3^{3}}
\no \\
F_{4} & = & g_{1a}\omega_{AdS_{3}} \wedge e^{a}+
 g_{2a}\;  \omega_{S_2^{3}} \wedge e^{a}+
 g_{3a}\;  \omega_{S_3^{3}} \wedge e^{a}
\eea
where  $\hat \omega_{AdS_{3}}$ and $\hat \omega_{S^{3}_{2,3}}$ are the volume 
forms on the unit-radius spaces $AdS_{3}$ and $S^{3}_{2,3}$ respectively, 
and  $e^{a}$, for $ a=1,2$  is an orthonormal frame on $\Sigma$. The volume forms 
of the full space-time metric are related to the ones with unit volume by 
$ \omega_{AdS_{3}} = f_1^3 \, \hat \omega_{AdS_{3}}$, and 
$\omega_{S_{2,3}^{3}} = f_{2,3}^3 \, \hat \omega_{S_{2,3}^{3}}$. In terms of an arbitrary 
system of local complex coordinates $w,\bar w$ on $\Sigma$, the metric 
$ds^2_\Sigma$ in (\ref{metricansatza}) reduces to the conformal form,
\bea
\label{rhofactor}
ds_{\Sigma}^2= 4 \rho^{2 } |dw|^2
\eea
The metric factors $f_1, f_2, f_3, \rho$, as well as the flux fields $b_1, b_2, b_3$, and 
$g_{1a}, g_{2a}$, and $g_{3a}$,  only depend on  $\Sigma$.
The Ansatz automatically respects $SO(2,2)\times SO(4)\times SO(4)$ symmetry,
which may also be viewed as the symmetry of an AdS/CFT dual 1+1-dimensional
conformal interface/defect in the 3-dimensional M2 brane CFT. 

\subsection{Equations for 1/2 BPS solutions}

In \cite{D'Hoker:2008wc}, the BPS equations governing half-BPS solutions 
were reduced to constructing a Riemann surface $\Sigma$ with boundary, 
a real positive harmonic function $h$ on $\Sigma$, and the solution to a first 
order partial differential equation on $\Sigma$ for a complex-valued field $G$, 
subject to a point-wise quadratic constraint.
The partial differential equation for $G$ is given by,
\bea
\label{gequation}
\p_w G = \half (G + \bar G) \p_w \ln h
\eea
for an arbitrary complex coordinate system $w, \bw$ on $\Sigma$.
For solutions which are locally asymptotic to $AdS_4 \times S^7$ 
(referred to as ``case I" in \cite{D'Hoker:2008wc}), 
the field $G$ is subject to the following point-wise quadratic constraint,
\bea
\label{gc}
|G(w, \bw)|^2 \geq 1 \qquad \qquad  {\rm for~all} ~  (w,\bw) \in \Sigma
\eea
We define the  following function $W$ on $\Sigma$,
\bea
W^2 \equiv 4 |G|^4 + (G - \bar G)^2
\eea
Assuming that the point-wise quadratic constraint $|G|^2 \geq 1$ is obeyed, 
we automatically have $W^2 \geq 0$, so that $W$ is real.
The special value $W^2=0$ corresponds to either $G=+i$ or $G=-i$.
The metric factors in (\ref{metricansatza}) are then expressed as follows,,
\bea
\label{metricfactors}
f_{1}^{6}
	&=&
	{ h^{2} W^{2} \over 16^2 (|G|^{2}-1)^{2}}
\no \\
f_{2}^{6}
	&=&
	 {h^{2} (|G|^{2}-1)\over 4\, W^{4}} \left ( 2 |G|^{2} + i (G-\bar G) \right )^{3}
\no \\
f_{3}^{6}
	&=&
	 {h^{2} (|G|^{2}-1)\over 4\, W^{4}} \left ( 2|G|^{2} - i (G-\bar G) \right )^{3}
\eea
The metric factor $\rho$ in (\ref{rhofactor}) is given by,
\bea
\rho^{6}
&=&
{\left | \partial_{w} h\right | ^6 \over 16^{2} h^{4}} \big(|G|^{2} -1 \big) W^{2}
\eea
The anti-symmetric tensor field-strengths are expressed in terms of $g_{1a}, g_{2a}$,
and $ g_{3a}$. They can be simply related to currents $\p_w b_1$, $\p_w b_2$, and 
$\p_w b_3$,  on $\Sigma$, which are conserved as a result of the Bianchi 
identities, and are given as follows,
\bea
\label{bi}
(f_1)^3 g_{1w} = - \p_w b_1
&=&  -{ 3 W^2 \p_w h \over 32 G (|G|^2 -1)}
- { 1+|G|^2 \over 16 G (|G|^2-1)^2} \, J_{w}
\no\\
(f_2)^3 g_{2w} = - \p_w b_2 
&=& -{ (G+i)\left  ( 2|G|^2 +i (G -\bar G) \right  )^2 \over W^4} \, J_w
\no\\
(f_3)^3 g_{3w} = - \p_w b_3 
&=& + { (G-i)\left  ( 2|G|^2 - i(G - \bar G) \right )^2 \over W^4} \, J_w
\eea
where the following intermediate quantity was used for notational compactness,
\bea
J_{w} = \half  (G \bar G - 3 \bar G^2 + 4 G \bar G^3) \p_w h + h G  \p_w \bar G
\eea
 It was shown in \cite{D'Hoker:2008wc} that the equations of motion of  as well as the Bianchi identities are satisfied for any harmonic $h$ and any $G$ which solves (\ref{gequation}).

\subsection{The $AdS_4 \times S^7$ solution}

The simplest solution is the maximally symmetric $AdS_4\times S^7$ itself.
The corresponding Riemann surface is the infinite strip,
\bea
\label{sigmadefi}
\Sigma=\{ w \in \bC, ~ w = x+iy, ~  x \in \bR, \; 0\leq y\leq \pi/2\}
\eea
Note that the Riemann surface $\Sigma$ has two boundary components, namely 
$y=0$ and $y = \pi/2$. In these coordinates, the functions  $h$ and $G$ for the 
$AdS_4 \times S^7$ solution are given by,\footnote{For simplicity, we shall exhibit 
the solutions with unit $AdS_4$-radius. The solution for general $AdS_4$-radius 
$R_0$ may be obtained by scaling $h \to R_0^3 h$, while leaving $\Sigma$, 
$w, \bar w$, and $G$ unchanged.  As a result, the metric factors scale as follows, 
$f_i \to R_0 f_i$,  $\rho \to R_0 \rho$, while the flux fields scale as $b_i \to R_0^3 b_i$ 
for $i=1,2,3$.}
\bea
\label{hgads}
h &=&
4i \big(\sh (2w) - \sh(2\bar w) \big)
\no\\
G &=&
i {\ch (w + \bar w) \over \ch (2 \bar w)}
\eea
It is easy to check that the partial differential equation (\ref{gequation}) as well as the 
point-wise quadratic constraint (\ref{gc}) are satisfied.
Using  (\ref{metricfactors}), the metric factors become,
\bea
\label{adssol}
f_1 =  \ch(2 x)
\qquad
f_2 = 2 \cos (y)
\qquad
f_3 = 2 \sin(y)
\qquad \qquad
\rho = 1
\eea
The boundary may be characterized by the vanishing of the harmonic function $h=0$,
or alternatively, by $G = \pm i $.
On the lower boundary of the strip $\Sigma$ where $y=0$ one has $G=+i$, which implies
that the radius $f_2$  of $S_2^3$ vanishes. On the upper boundary of the strip $\Sigma$
where $y=\pi/2$ one has  $G=-i$, which implies that the radius $f_3$ of $S_3^3$
vanishes.  The boundary of $AdS_4$, on the other hand, is located at 
$x = \pm \infty$.

\subsection{The half-BPS Janus solution
\label{twopolestrip}}

It turns out that there is a simple deformation of $AdS_{4}\times S^{7}$ 
(reviewed in the previous section) which corresponds to a Janus solution in M-theory. 
As before, the holomorphic coordinate is denoted as $w=x+i y$ and the strip is 
parametrized as (\ref{sigmadefi}). The harmonic function $h$ is 
taken to be proportional to the one of (\ref{hgads}), but is rescaled in order for  
the asymptotic $AdS_{4}\times S^{7}$ solution  to  have the same radius  as the 
undeformed solution (\ref{hgads}), namely with unit $AdS_4$-radius,
\be
 h = {4i\over \sqrt{1+\lambda^{2}}} \Big ( \sh (2w) - \sh(2\bar w) \Big )
\ee
The expression for $G$ is given by
\bea
G = i { \ch (w+\bw) + \lambda \sh (w - \bw) \over \ch (2 \bw)}
\eea
It is easy to check that the partial differential equation (\ref{gequation}) 
as well as the positivity constraint (\ref{gc}) are satisfied for any real value of $\lambda$. 
The solution forms a one-parameter deformation of $AdS_{4}\times S^{7}$, which 
one recovers by setting $\lambda=0$. 

\sm

The metric factors can be expressed concisely in terms of two functions,
\bea
F_{+}(x,y)
&=& 
1+ 2 \lambda \Big (  \sh (2x)  +  \lambda \Big ) \cos^{2} (y)  /\ch^{2} (2x) 
\no \\
F_{-}(x,y)
&=&
1-2\lambda \Big (  \sh (2x)  -  \lambda \Big ) \sin^{2} (y)  /\ch^{2} (2x)
\eea
The metric factors (\ref{metricfactors}) become,
\bea
\label{metfacsol}
f_{1} 
&=& 
{\ch(2x) \over \sqrt{1+\lambda^{2}}} F_{+}(x,y)^{1\over 6}  F_{-}(x,y)^{1\over 6}  
\no \\
f_{2}
&=& 
2 \cos (y) \, F_{+} (x,y)^{1\over 6} F_{-}(x,y)^{-{1\over 3}} 
\no \\
f_{3}
&=& 
2 \sin ( y ) \,  F_{-}(x,y) ^{1\over 6} F_{+}(x,y)^{-{1\over 3}} 
\no \\
\rho
&=&  
F_{+}(x,y)^{1\over 6} F_{-}(x,y)^{1\over 6}
\eea

\begin{figure}\label{fig1}
\centering
\includegraphics[ scale=0.50]{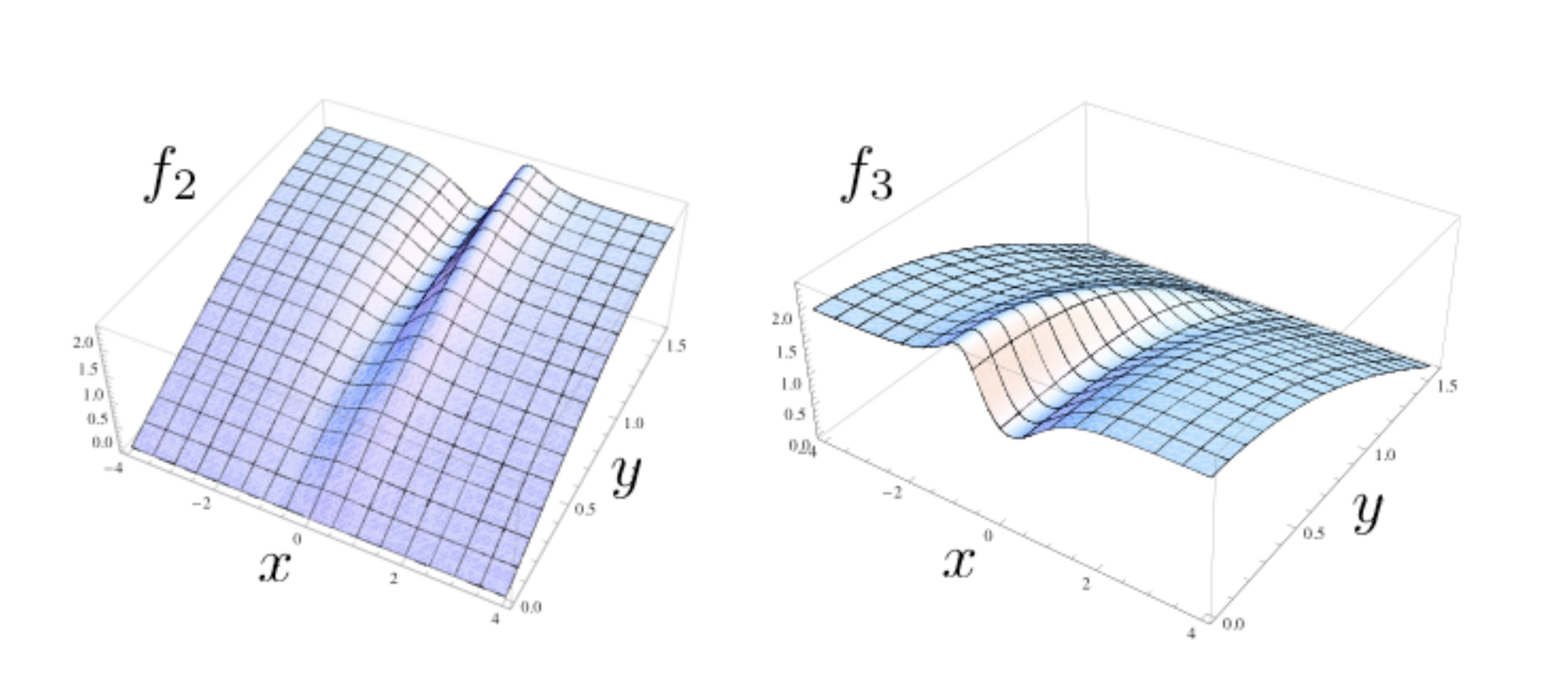}
\caption{ Metric factors of the two spheres  for the deformation $\lambda=1$ .}
\end{figure}

Note that the metric factor for the base space $\Sigma$ becomes non-trivial for 
the deformed solution. Another  interesting feature of this solution is that the size 
of each three sphere at the part of the boundary where it does not vanish, i.e. $y=0$ 
for $f_{2}$ and $y=\pi/2$ for $f_{3}$, becomes $x$ dependent in contrast to the 
$AdS_{4}\times S^{7}$ solution where it is constant.

\begin{figure}\label{fig2}
\centering
\includegraphics[ scale=0.50]{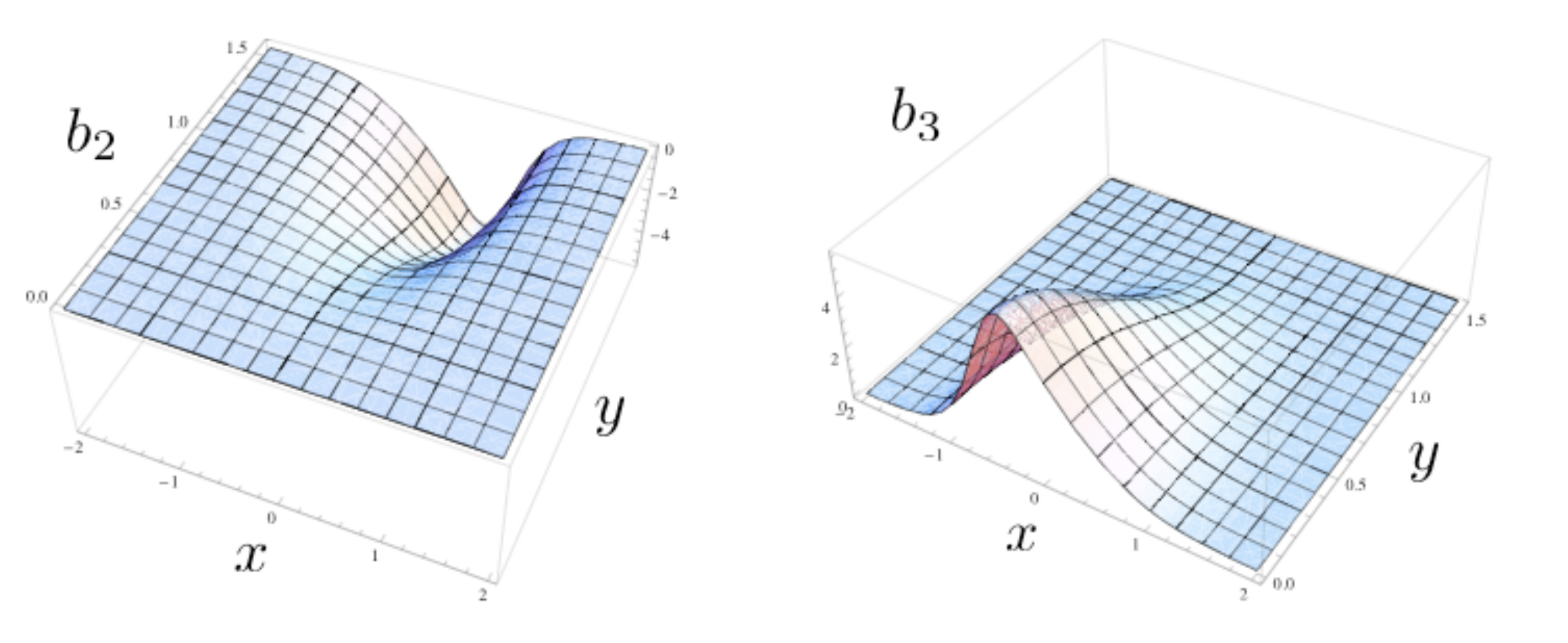}
\caption{AST potentials along the two three spheres for the deformation $\lambda=1$ .}
\end{figure}

The AST fields along the two three spheres  are non-vanishing for the deformed solution.  
The formulae for the fieldstrengths (\ref{bi}) can be integrated and one obtains the following expressions for the AST potentials.
\bea
\label{bsol}
b_{2}&=& - { 8  \lambda\sqrt{1+\lambda^{2}}\sin^{4} (y) \over  \ch^{2}(2x)\;  F_{-}(x,y)}
\no \\
b_{3}&=&   { 8  \lambda\sqrt{1+\lambda^{2}}\cos^{4}( y )\over  \ch^{2}(2x)\;  F_{+}(x,y)}
\eea
The explicit formulae for the AST potential $b_{1}$ are also easy to calculate but will not be needed in the following. We have checked that the solutions indeed satisfies all equations of motion of 11-dimensional supergravity. The solution preserves sixteen of the thirty two supersymmetries by construction  \cite{D'Hoker:2008wc}.

\subsection{Holographic interpretation}

In the previous section we utilized the coordinate system in which $\Sigma$ is 
a strip  parametrized as $\Sigma=\{(x,y), x \in \bR, \; 0\leq y\leq \pi/2\}$. 
The boundaries of $\Sigma$ at $y=0,\pi/2$ are the location where the volume 
of the first and second three sphere vanishes. For fixed $x$ the finite $y$-interval 
together 
with the two three spheres produces a deformed seven sphere with $SO(4)\times SO(4)$ 
symmetry. The asymptotic $AdS$ region is reached by taking $x\to \pm \infty$.  
In this limit the leading terms of the metric factors (\ref{metfacsol})  behave as follows,
\be
\label{limmet}
ds^{2} = 
{1\over 4} e^{\pm 4 x} \; {-dt^{2}+ds^{2}+dz^{2}\over z^{2}} 
+ 4 \sin^{2} (y) \;ds_{S_2^3}^2 
+ 4 \cos^{2 } (y)\; ds_{S_3^3}^2 
+  4(dx^{2}+dy^{2}) +o( e^{\mp 4 x})
\ee
which is the asymptotic form of $AdS_{4}\times S^{7}$. The leading terms for the 3-form 
potential $C_3$ in the limit $x\to \pm \infty$ can be obtained from (\ref{bsol}),
\bea
\label{expthreef}
C_3
= {1\over 16} e^{\pm 6x}\hat \omega _{AdS_3} 
-{32\lambda\over \sqrt{1+\lambda^{2}} }e^{\mp 4 x} 
\Big( \sin^{4} (y) \; \hat \omega _{S_2^3} 
- \cos^{4} (y) \; \hat \omega _{S_3^3} \Big)
 + o(e^{\mp 2 x})
\eea
The first term in (\ref{expthreef}) produces the four form flux supporting the 
$AdS_{4}\times S^{7}$ solution. The second term vanishes when  
$\lambda=0$, which corresponds to the undeformed solution.  Note that the deformed solution  has a nontrivial profile for the AST potential  along the spheres, but it is easy to see that  the conserved M5-brane charge integrates to zero.

\sm

To make contact with the discussion of section 2, we first note that the dependence 
on the $y$ and sphere coordinates in the second and third terms  in  (\ref{expthreef}) 
corresponds to a specific  AST spherical harmonic on the sphere. The Kaluza-Klein 
reduction \cite{Biran:1983iy} produces the pseudo-scalar field $\phi$ of mass $m^{2}=-2$ transforming in the ${\bf 35_{c}}$ of the $SO(8)$ R-symmetry. 

\sm

We can relate the coordinate $x$ which parametrizes the $AdS_{3}$ slicing 
coordinate in the strip parametrization to the coordinate $\mu$ which was 
employed in section 2 by,
\be
\mu \mp \pi/2 = e^{\mp 2x}
\ee
valid in the limit $x\to \pm \infty$. It follows from (\ref{expthreef}) that the 
pseudo-scalar field mode associated with the KK reduction of the AST 
potential has the following behavior,
\be
\lim_{\mu \to \pm \pi/2 }\phi = {\rm constant} \times (\mu \mp \pi/2 )^{2}+ o(\mu \mp \pi/2)^{4}
\ee
Consequently, the holographic behavior of $\phi$  is exactly of the type discussed in section 2. 
The holographic interpretation of our solution is that a dimension two operator which 
preserved $SO(4)\times SO(4)$ R-symmetry, as well as sixteen supersymmetries, is 
sourced at a 1+1-dimensional interface/defect. 

\sm

In the context of BPS Janus solutions in Type IIB the related CFT analysis of 
supersymmetric interface/defects in $\cN=4$ super Yang Mills was carried out  
in \cite{Clark:2004sb,D'Hoker:2006uv,Gaiotto:2008sd,Gaiotto:2008sa}. 
It would be interesting to pursue a similar analysis for the M2-brane CFT, 
where significant progress in the formulation of the theory has been made 
recently \cite{Bagger:2006sk,Bagger:2007jr,Gustavsson:2007vu}. 
See \cite{Honma:2008un,Ryang:2009bs} for other attempts to obtain 
Janus like solutions for the M2-brane worldvolume theory.

%%%%%%%%%%%%%%%%%%%%%%%%%%%%%
%%%%%%%%%%%%%%%%%%%%%%%%%%%%%
\section{Janus solutions in ABJM theory}
\setcounter{equation}{0}
%%%%%%%%%%%%%%%%%%%%%%%%%%%%%
%%%%%%%%%%%%%%%%%%%%%%%%%%%%%

Recently, ABJM \cite{Aharony:2008ug} 
found a new AdS/CFT correspondence between certain quotients of 
$AdS_{4}\times S^{7}$ in M-theory, and 2+1-dimensional CFTs which preserve $\cN=6$ 
superconformal symmetry.  The remarkable benefit of this correspondence lies
in the fact that it provides a standard field theory description of the corresponding
2+1-dimensional CFTs.  In this section, we show that for each of the ABJM quotients,
there exists a corresponding regular ABJM Janus solution, which may be obtained 
by quotienting the M-theory Janus solution of the previous section \`a la ABJM.

\subsection{Construction of the ABJM Janus solution}

The supergravity description of ABJM theory is given by the quotient 
$AdS_{4}\times S^{7}/Z_k$ of the maximally symmetric vacuum $AdS_4 \times S^7$ 
by  the cyclic group $Z_k$ for $k \geq 1$. Here, $Z_k$ acts on $S^7$, 
but does not act on $AdS_4$. 
The action of $Z_k$ on $S^7$ has no fixed points, and the resulting quotient
may be viewed as a line bundle over $CP_3$ whose fiber $S^1$ has radius $1/k$.
The quotient breaks the $SO(8)$ R-symmetry to $SU(4)\times U(1)$.

\sm

The action on $S^7$ may be rendered more
explicit by embedding $S^{7}$ into $\bR^{8}$, and parametrizing $\bR^8$ by 
four complex coordinates $z_{i}$, with $i=1,2,3,4$. The $Z_{k}$ transformation 
is then defined by $z_{i}\to z_{i} \, e^{2\pi i / k}$ for any integer $k \geq 1$, and the 
quotient is obtained by identifying the points $z_{i} $ and $z_{i} \, e^{2\pi i / k}$
for all four directions $i=1,2,3,4$.  

\sm

To construct the ABJM Janus solution, it will be more useful 
to exhibit the $Z_k$ transformations in terms of real coordinates, on which
the action of the full $SO(8)$ is manifest. The corresponding representation 
matrix $\gamma$ in the group $ SO(8)$ is then given by $\g = e^{ 2 \pi H/k}$,
where $H$ is the following generator of the Lie algebra $SO(8)$, 
\bea
H = \left ( \matrix{ H_2 & 0 \cr 0 & H_3 \cr} \right )
\hskip 0.8in 
H_{2}=H_{3} = \left ( \matrix{\ep & 0 \cr 0 & \ep \cr} \right )
\hskip 0.8in 
\ep = \left ( \matrix{0 & -1 \cr 1 & 0 \cr} \right )
\eea
The notation $H_2$ and $H_3$ has been introduced for the following reason. 
The M-theory Janus solution is characterized by the breaking of the Lie algebra 
$SO(8)$ to $SO(4)_2 \oplus SO(4)_3$, the two $SO(4)_i$ subalgebras being the 
isometries of the spheres $S_i^3$ for $i=2,3$ of the Janus solution in (\ref{metricansatza}). 
In the above partition of $H$ into the direct sum of $H_2$ and $H_3$,
the generators are arranged so that $H_2 \in SO(4)_2$, and $H_3 \in SO(4)_3$.
As a result, the action of $Z_k$ on $S^7$ descends to the Janus solution 
as a $Z_k$ action on $S_2^3$ and $S_3^3$, whose explicit form is given by
the $SO(4)$ Lie group matrices $\g_2 = e^{2 \pi H_2/k}$ and $\g_3 = e^{2 \pi H_3/k}$.

\sm

The action of $\g_{2,3}$  on $S^3_{2,3}$ is again without fixed points.
Furthermore, the action of $\g_{2,3}$ leaves all the other ingredients of
the Ansatz of (\ref{metricansatza}), (\ref{fluxansatz}), (\ref{rhofactor}),
and of the solution functions $h$ and $G$ invariant.  Thus, we are 
guaranteed that the quotient of the M-theory Janus solution by the
action of $Z_k$ will produce a regular family of solutions, parametrized 
by the same parameter $\lambda$ as the M-theory Janus was. The only 
change to the geometry resides in the quotient of $S_2^3 \times S_3^3$ by $Z_k$.

\sm

The quotienting of $S^3$ by $Z_k$ reduces the isometry group from 
$SO(4)$ to an $SU(2) \times U(1)$ subgroup. Thus, we expect the ABJM
Janus solution to have a compact bosonic symmetry group 
$SU(2) \times SU(2) \times U(1)^2$, as well as, of course,  the full 
isometry $SO(2,2)$ of $AdS_3$, producing a total bosonic symmetry 
group
\bea
SO(2,2) \times SO(4) \times U(1)^2
\eea
Supersymmetry is also reduced under quotienting by $Z_k$, $k \not= 1,2$.
This reduction is entirely due to the reduction of the number of Killing
spinors on $S_2^3 \times S_3^3$, and proceeds in parallel to the 
corresponding reduction on $S^7$. The 4 independent Killing spinors 
on $S_2 ^3 \times S_3^3$ are reduced to only 3 Killing spinors 
on $(S_2 ^3 \times S_3^3)/Z_k $.
For $AdS_4 \times S^7$, this effect reduces the total number of supersymmetries 
from 32 to 24, while for the M-theory Janus solution, it reduces the number
of supersymmetries from 16 to 12. Given these bosonic and supersymmetries, 
the corresponding invariance superalgebra of the 
ABJM Janus solution is readily obtained,
\bea
OSp(3|2,\bR) \times OSp(3|2,\bR) 
\eea
which is a subgroup of the $OSp(4|2,\bR) \times OSp(4|2,\bR)$ algebra of
the M-theory Janus solution.

\subsection{Structure of the supergravity multiplets}

As pointed out in \cite{Nilsson:1984bj,Halyo:1998pn}, the  spectrum  of the KK 
reduced theory can be obtained by decomposing the KK spectrum of 
$AdS_{4}\times S^{7}$ with respect to  $SU(4)\times U(1)$ and keeping 
only the zero charge sector of the $U(1)$. For the  fields of $N=8$ gauged supergravity  
in Table 2, this implies the following decompositions of $SO(8)$
representations under its $SU(4) \times U(1)$ subgroup, 
 \bea
 \label{quotable}
 {\bf 28_{v}} & \to & {\bf 1_{0}} + {\bf 6_{2}} + {\bf 6_{-2}} + {\bf 15_{0}} 
 \no\\
 {\bf 35_{v}} &\to& {\bf 10_{2}} + {\bf \overline{10}_{-2}} + {\bf 15_{0}}
 \no\\
{\bf 35_{c}} &\to& {\bf 10_{2}} + {\bf \overline {10}_{-2}} + {\bf 15_{0}}
 \eea
As a result,  out of the  35 pseudo-scalar $0^{- }$ fields transforming in the 
${\bf 35_{c}}$ of $SO(8)$  in the KK spectrum, fifteen survive the quotient 
and transform in the ${\bf 15}$ of $SU(4)$. These fields  have mass 
$m^{2}=-2$ and are dual to dimension 2 operators.  The linearized analysis 
of section one is expected to apply to these states, and it provides further
evidence for the existence of Janus-like interface/defect solutions in this theory.

\sm

It is possible to characterize the dual operator using the ABJM worldvolume theory.  
The scalars and fermion fields transform as bi-fundamentals of a $U(N)\times U(N)$ 
gauge theory. There is an SU(2) doublet of bosons $A_{1},A_{2}$  and fermions 
$\lambda_{1},\lambda_{2}$ which transform as $(N,\bar N)$ under 
$U(N)\times U(N)$ gauge symmetry, whereas a second pair of bosons 
$B_{1},B_{2}$ and fermions $\chi_{1},\chi_{2}$  transforms as  $(\bar N,N)$. 
The boson  and fermion fields can be assembled into the following multiplets
\cite{Aharony:2008ug,Benna:2008zy, Hosomichi:2008jb},
\be
Y^{A}=(A_{1},A_{2},B_{1}^{\dagger},B_{2}^{\dagger})
\hskip 1in  
\Psi^{A}=(\lambda_{1},\lambda_{2},\chi_{1}^{\dagger},\chi_{2}^{\dagger})
\ee
which transform in the ${\bf 4}$ representation of $SU(4)$, wheras the conjugate 
fields  $(Y^{A})^{\dagger}=Y_{A}$ and  $(\Psi^{A})^{\dagger}=\Psi_{A}$ transform 
in the ${\bf \bar 4}$ of $SU(4)$.
The following operators
\be
O_{1}= \tr \Big ( Y^{A}Y_{B}- {1\over 4} \delta^{A}_{\;B} Y^{C}Y_{C} \Big )
\hskip 1in  
O_{2}= \tr \Big ( \Psi^{A}\Psi_{B}- {1\over 4} \delta^{A}_{\;B} {\Psi}^{C}\Psi_{C} \Big )
\ee
are conformal primary operator of dimensions $\Delta=1$ and $\Delta=2$ 
respectively which transform as the ${\bf 15}$ of $SU(4)$. Therefore, they can 
be identified  with the surviving scalar and pseudo-scalars in the 
quotient (\ref{quotable}). Applying the linearized analysis of section 2 to  
the field dual to one $O_{2}$ one expects that a Janus-like interface/defect 
solution exists for the $AdS_{4}\times S^{7}$ quotient.

%%%%%%%%%%%%%%%%%%%%%%%%%%%%%
\section{Discussion}
\setcounter{equation}{0}
%%%%%%%%%%%%%%%%%%%%%%%%%%%%%

In this paper we have presented exact solutions of 11-dimensional supergravity 
which are holographically dual to inserting a dimension 2 operator along a 
1+1-dimensional interface/defect in the maximally supersymmetric
2+1-dimensional CFT.  The M-theory 
Janus solution preserves 16 supersymmetries and has $SO(2,2)\times SO(4)\times SO(4)$ isometry group, while the ABJM Janus solution preserves 12 supersymmetries 
and has $SO(2,2) \times SO(4) \times U(1)^2$ isometry group. 
The symmetries combine  into an $OSp(4|2,R)\otimes OSp(4|2,R)$ invariance 
superalgebra of the M-theory Janus solution, and an $OSp(3|2,R)\otimes OSp(3|2,R)$
invariance superalgebra for the ABJM Janus solution. Both are subgroups of
the supergroup $OSp(8|4,R)$ of the $AdS_{4}\times S^{7}$ vacuum \cite{D'Hoker:2008ix}. 
These solutions are analogs in M-theory of the Janus solution of Type IIB, even though 
no dilaton is present in M-theory.

\sm

There are several interesting open questions and directions for further research. 
\begin{enumerate}
\item 
Can the 11-dimensional Janus solutions be expressed as solutions
solely of the massless multiplet of the $\cN=8$ gauged supergravity.
\item 
Exact solutions, such as the M-theory and ABJM Janus solutions obtained here,
may be used to calculate interesting quantities using the machinery of AdS/CFT. 
For example, application of the methods developed in 
\cite{Clark:2004sb,Aharony:2003qf,Papadimitriou:2004rz} could be used to calculate correlation 
functions in the presence of the interface/defect.
\item 
As the ABJM theory enjoys a well-understood field theoretic CFT description, 
one may classify the possible interface/defect terms along the lines of 
\cite{D'Hoker:2006uv} and \cite{Gaiotto:2008sd}, establish their symmetries,
and derive the associated supergravity solutions. 
\item 
In Type IIB theory, the simplest Janus solution breaks all 
supersymmetries, has  a non-trivial dilaton profile, and vanishing 3-form flux fields. 
A natural question is whether M-theory Janus solutions exist with no, or further
reduced supersymmetry, and whether the corresponding supergravity solutions
lend themselves to exact construction.
\item 
Finally, does Type IIB supergravity support Janus-type solutions whose 
dilaton is constant, but whose 3-form and 5-form fields vary spatially ? 
The corresponding CFT dual would be $\cN=4$ super-Yang-Mills with identical 
gauge couplings on both sides of the interface/defect, and dimension 3 operators 
localized on this defect/interface. In other words, does Type IIB admit 
the Janus-type solutions characteristic of M-theory~?
\end{enumerate}
We plan to address some of these questions in the near future.

\bigskip\bigskip

\noindent{\Large \bf Acknowledgements}

\bigskip

We would like to thank Per Kraus for useful conversations.
MG gratefully acknowledges the hospitality of   the Department of Physics 
and Astronomy, Johns Hopkins University during the course of this work.

\end{document}